\documentclass{article}
\usepackage{amssymb,latexsym}
\usepackage[dvipdf]{epsfig}
\usepackage[dvipdf]{epsfig}
\def\be{\begin{equation}}
\def\ee#1{\label{#1}\end{equation}}

\newcommand{\ben}{\begin{eqnarray}}
\newcommand{\een}{\end{eqnarray}}
\def\wo{\widetilde{\omega}}
\def\lb{\label}

\begin{document}

\title{Fermion fields in Einstein-Cartan theory and the accelerated-decelerated transition in a primordial Universe}

\author{Marlos O. Ribas$^1$ and Gilberto M. Kremer$^2$\\
$^1$ Faculdades Integradas Esp\'{\i}rita, \\Rua Tobias de
Macedo Jr.  333, 82010-340 Curitiba, Brazil\\
 $^2$Departamento de F\'{\i}sica, Universidade Federal do
Paran\'a,\\ Caixa Postal 19044, 81531-990 Curitiba, Brazil}
\date{}
\maketitle

\begin{abstract}
In this work the accelerated-decelerated transition in a primordial Universe is investigated by using the dynamics of fermion fields within the context of Einstein-Cartan theory, where apart from the curvature the space-time is also described by a torsion field.
The model analyzed here has only a fermion field as the source of the gravitational field. The term associated with the spin  of the fermion field plays the role of the inflaton  which contributes to an accelerated regime whereas the one related to the fermion mass behaves as a matter field and is the responsible for a decelerated regime. Hence, by taking into account the spin of a massive fermion field it is possible to characterize the transition from the accelerated to the decelerated periods of the primordial Universe.
\end{abstract}


%%%%%%%%%%%%%%%%%%%%%%%%%%%%%%%%%%%%%%%%%%%%%%%
\section{Introduction}
%%%%%%%%%%%%%%%%%%%%%%%%%%%%%%%%%%%%%%%%%%%%%%%

The standard hot Big-Bang model is a solid theory which has predicted several essential observational successes among others the expansion of the Universe, its age, the origin of the cosmic background radiation and  the nucleosynthesis of the light elements \cite{peb}.

However there are other observational keys that the hot Big-Bang model  could not  explain like the problems of flatness,  horizon and unwanted relics \cite{linde}.
To overcome  these problems it has been assumed that after Planck's era the Universe has suffered a huge accelerated expansion in a very small time interval, which constitutes the inflationary period. Normally for the description of  the accelerated era  it has been used scalar fields -- the so-called inflaton -- with appropriate potentials which can solve the problems of the standard model \cite{guth1,guth2,guth3}.
Another possibility to generate accelerated regimes is to use the dynamics of fermions fields with suitable auto-interaction potentials, where the fermion field plays the role of the inflaton or dark energy \cite{gdm1,gdm2,gdm3,gdm4,gdm5}.

In some works \cite{gas1,gas2,cris} it was suggested that the inflation could be derived by the spin density which was present in the primordial period of the Universe. In this case the spin density acts as a repulsive source and its rapid decay contributes to the small time interval of the accelerated expansion. It has been also suggested that cosmological observations could detect small remnant of the inflationary period generated by the  torsion effects \cite{garcia} which could also solve the problem of the sign of the cosmological constant \cite{bor}. The preliminaries works \cite{ko,k} on cosmological models involving  space-time torsion considered fluids with rotational symmetries as  sources of the gravitational field. Recently, some cosmological models were proposed which takes into account the effects of the torsion in order to describe the present accelerated expansion of the Universe \cite{capo1,capo2,min}. Within  this context it has also been studied eigenspinors of the  charge conjugation operator -- the so-called dark spinors -- as source of the torsion \cite{bo}. These spinors could describe the inflationary regime as well as the present accelerated expansion. The theories of  space-time with torsion  appear naturally in super-symmetric gravitational models, hence  a possible extension of general relativity should include the torsion effects due to spin sources.

In this work the dynamics of fermion fields which takes into account the spin density is used in order to exploit the inflationary regime. The spin density plays the role of the inflaton and is responsible for the accelerated expansion of the primordial Universe. It differs from our previous works \cite{gdm1,gdm2} since the self-interaction potential of the fermion field is not taken into account. To include the spin into the general gravitational theory it is necessary to work within the framework of the  Einstein-Cartan theory, where as usual the matter-energy is a  source of the space-time curvature and the spin density is related with the torsion of the space-time manifold \cite{hell1,hell2,hell3,pon,watanabi,pereira,ace,wa}.

This work is organized as follows. In section 2 it is introduced the main properties of the Einstein-Cartan theory which are used to obtain the Einstein and Dirac equations in section 3. The field equations in a Robertson-Walker metric are derived in section 4 and the inflationary solution related with the proposed model is presented in section 5. In section 6 the main conclusions of the work are introduced.
The metric signature $(+, -, -, -)$  is adopted and natural units $c=\hbar=1$ are used. Furthermore, parenthesis and brackets enclosing two indices denote symmetrization and anti-symmetrization with respect to the indices, respectively.

%%%%%%%%%%%%%%%%%%%%%%%%%%%%%%%%%%%%%%%%%%%%%%%
\section{Preliminaries}
%%%%%%%%%%%%%%%%%%%%%%%%%%%%%%%%%%%%%%%%%%%%%%%

 In the context of Einstein-Cartan theory the affine connection $\Gamma^\rho_{\mu\nu}$ is not symmetric and
 its anti-symmetric part defines the torsion tensor \cite{watanabi}
 \be
 {C^{\lambda}}_{\mu\nu}=\Gamma^{\lambda}_{\mu\nu}-\Gamma^{\lambda}_{\nu\mu}\equiv2\Gamma^{\lambda}_{[\mu\nu]}.
 \ee{1}

 The relationship of the affine connection $\Gamma^\lambda_{\nu\mu}$ with the Cristoffel symbol $\widetilde\Gamma^\lambda_{\nu\mu}$ is obtained via the metricity condition $\nabla_{\sigma}g_{\mu\nu}=0,$ yielding
  \be
  \Gamma^\lambda_{\nu\mu}=\widetilde\Gamma^\lambda_{\nu\mu}+{K^{\lambda}}_{\nu\mu},
  \ee{2}
  where ${K^{\lambda}}_{\nu\mu}$ is the contortion tensor defined by
  \be
  {K^{\lambda}}_{\nu\mu}=\frac{1}{2}\left({C^{\lambda}}_{\nu\mu}+{C_{\nu\mu}}^{\lambda}
  +{C_{\mu\nu}}^{\lambda}\right).
  \ee{3}

When one is dealing with fermion fields as  sources of the gravitational field, it is convenient to use the tetrad formalism. The tetrad $e^a_\mu$ defines an orthonormal set of vectors  that satisfies the relationship
$g_{\mu\nu}=e^a_\mu e^b_\nu\eta_{ab}$, where $\eta_{ab}$ is the Minkowski metric tensor.  In the tetrad formalism  the metricity condition corresponds to the tetrad condition which reads
\be
{\cal D}_\nu e^{a\mu}\equiv \partial_{\nu}e^{a\mu}+\Gamma^{\mu}_{\rho\nu}e^{a\rho}+{\omega_{\nu}}^{ab}e_b^\mu=0,
\ee{4}
 where ${\omega_{\nu}}^{ab}$ denotes the spin connection.

 In terms of the spin connections the Riemann tensor becomes
\be
{R^{ab}}_{\mu\nu}=\partial_\mu{\omega_{\nu}}^{ab}-\partial_\nu{\omega_{\mu}}^{ab}
+{\omega_{\mu}}^{ac}{\omega_{\nu c}}^{b}-{\omega_{\nu}}^{ac}{\omega_{\mu c}}^{b},
\ee{5}
and the the Ricci tensor is given by $R_{\mu\nu}=e_a^\sigma e_{b\nu}{R^{ab}}_{\sigma\mu}$.

From the tetrad condition  (\ref{4}) one can write
\be
{\omega_\mu}^{ab}={\wo_\mu}^{\;\; ab}+{K^{ab}}_\mu,
\ee{6}
 due to equation (\ref{2}) and the definition of the Cristoffel symbol $\widetilde\Gamma^\lambda_{\nu\mu}$. The spin connection ${\wo_\mu}^{\;\; ab}$ is only a function of the tetrad, i.e.,
\ben\lb{7}
{\wo_\mu}^{\;\; ab}={e^{a\rho}\over2}\left(\partial_\mu e^b_\rho-\partial_\rho e^b_\mu\right)-
{e^{b\rho}\over2}\left(\partial_\mu e^a_\rho-\partial_\rho e^a_\mu\right)+{e^{a\rho}\over2}\left(\partial_\sigma e^c_\rho-\partial_\rho e^c_\sigma\right)e^{b\sigma}e_{c\mu}.
\een

The Ricci tensor $R_{\mu\nu}$ can also be written in terms of a part which does not depend on the contortion tensor ${K^{\lambda}}_{\mu\nu}$ and another which is a function of it, namely
\ben\lb{8}
R_{\mu\nu}=\widetilde R_{\mu\nu}+\widetilde\nabla_{\lambda}{K^{\lambda}}_{\mu\nu}
-\widetilde\nabla_{\nu}{K^{\lambda}}_{\mu\lambda}+{K^{\lambda}}_{\theta\lambda}{K^{\theta}}_{\mu\nu}
-{K^{\lambda}}_{\theta\nu}{K^{\theta}}_{\mu\lambda},
\een
where the  $\widetilde R_{\mu\nu}$ and $\widetilde\nabla_{\lambda}$ are the Ricci tensor and the covariant derivative referred to the Cristoffel symbol $\widetilde\Gamma^\lambda_{\nu\mu}$, respectively.

The Dirac matrices in the Minkowski space-time are denoted by  $\gamma_{a}$ with $a=0,1,2,3$ whereas the generalized Dirac matrices -- given by $\gamma^\mu=e^{a\mu}\gamma_a$ -- satisfy the Clifford algebra $\{\gamma^\mu, \gamma^\nu\} =2 g^{\mu\nu}$, where the braces denote the anti-commutation relation. As usual,  the brackets  $\left[\gamma_a,\gamma_b\right]$ will denote the commutation relation and $\gamma^5=-\imath \, \gamma^0\,\gamma^1\,\gamma^2\,\gamma^3$.

%%%%%%%%%%%%%%%%%%%%%%%%%%%%%%%%%%%%%%%%%%%%%%%
\section{Einstein and Dirac equations}
%%%%%%%%%%%%%%%%%%%%%%%%%%%%%%%%%%%%%%%%%%%%%%%

The Lagrangian density of a fermion field is given by
\be
{\cal L}=\frac{\imath}{2}\left(\overline\psi\gamma^\mu D_\mu\psi
-\overline D_\mu\overline\psi\gamma^\mu\psi\right)-m\overline\psi\psi-V,
\ee{9a}
where  $\psi$ and $\overline\psi=\psi^\dag\gamma^0$ denote the spinor field and its adjoint, respectively, $m$ is the fermion mass and  $V$  its potential of self-interaction. The covariant derivatives  $D_\mu\psi$ and $\overline D_\mu\overline\psi$ are given
in terms of the spin connection ${\omega_\mu}^{ab}$ by
\ben\lb{10}
D_\mu\psi=\partial_\mu\psi+\frac{1}{8}{\omega_\mu}^{ab}\left[\gamma_a,\gamma_b\right]\psi,
\qquad
\overline D_\mu\overline\psi=\partial_\mu\overline\psi
-\frac{1}{8}{\omega_\mu}^{ab}\overline\psi\left[\gamma_a,\gamma_b\right].
\een

 Within the framework of  Einstein-Cartan theory, the action for a fermion field minimally coupled to the gravitational field reads
\ben\lb{9}
S=\int e\,\Biggl[-\frac{1}{16\pi G}e^\mu_ae^\nu_b{R^{ab}}_{\mu\nu}
+\frac{\imath}{2}\Bigl(\overline\psi\gamma^\mu D_\mu\psi
-\overline D_\mu\overline\psi\gamma^\mu\psi\Bigr)-m\overline\psi\psi-V\Biggr]d^4x,
\een
where $e=\det(e^\mu_a)$ and $G$ is the gravitational constant.

The field equations are obtained from the action (\ref{9}) as follows. First the variation of the action (\ref{9}) with respect to the tetrad leads to Einstein field equations
\be
R_{\mu\nu}-\frac{1}{2}R g_{\mu\nu}=8\pi G\, T_{\mu\nu},
\ee{12}
where the  energy-momentum tensor is   given by
\be
T_{\mu\nu}=
{\imath\over2}\left(\overline\psi\gamma_\nu D_\mu\psi- {\overline D}_\mu\overline\psi\gamma_\nu\psi\right)
-{\cal L}g_{\mu\nu}.
\ee{13}

Next, Dirac's equations follow from the variation of the action (\ref{9}) with respect to  $\overline\psi$ and $\psi$,
yielding
\be
\imath\gamma^\mu D_\mu\psi-m\psi-\frac{dV}{d\overline\psi}=0, \qquad \imath\overline D_\mu\overline\psi\gamma^\mu+m\overline\psi+\frac{dV}{d\psi}=0.
\ee{14}

Finally, the variation of the action (\ref{9}) with respect to the spin connection implies the following expression for the torsion tensor
\be
{C^{\mu }}_{\kappa\lambda}=-4\pi G\,\epsilon_{abcd}\,e^{a}_{\lambda} e^b_\kappa e^{c\mu}\left(\overline\psi\gamma_5\gamma^d\psi\right),
\ee{15}
where  $\epsilon_{abcd}$ is the Levi-Civita tensor.

Once the torsion tensor is given by equation (\ref{15}) one can obtain from (\ref{3}) the following expression for the contortion tensor:
\be
{K^{\lambda }}_{\nu\mu}=-2\pi G\,\epsilon_{abcd} \,e^{a}_\mu e^b_\nu e^{c\lambda} \left(\overline\psi\gamma_5\gamma^d\psi\right).
\ee{16}

The Einstein field equations (\ref{12}) may  be rewritten by introducing the symmetric Ricci tensor $\widetilde R_{\mu\nu}$ through the use of the relationship (\ref{8}).  Indeed, from the resulting equation one may obtain two equations, one referring to its symmetric part and another to its antisymmetric part. The symmetric part is similar to Einstein's field equations and reads
 \be
\widetilde R_{\mu\nu}-\frac{1}{2} \widetilde R g_{\mu\nu}=8\pi G\,\left( \widetilde T_{\mu\nu}-\frac{3}{2}\pi G g_{\mu\nu}\sigma^2\right),
\ee{17}
where it was introduced the abbreviation $\sigma^2=(\overline\psi\gamma_5\gamma_d\psi)^2$ and the symmetric energy-momentum tensor of the sources is given by
\ben\nonumber
 \widetilde T_{\mu\nu}=
{\imath\over2}\left(\overline\psi\gamma_{(\nu} \widetilde D_{\mu)}\psi-\widetilde {\overline D}_{(\mu}\overline\psi\gamma_{\nu)}\psi\right)-\widetilde{\cal L}g_{\mu\nu}.
\een
Above, $\widetilde{\cal L}$ is the Lagrangian without the torsion effects, the derivatives $\widetilde D_{\mu}$ and $\widetilde {\overline D}_{\mu}$ are given by the expressions (\ref{10}) by replacing the spin connection ${\omega_\mu}^{ab}$ by ${\widetilde\omega_\mu}^{\;\; ab}$.

The corresponding antisymmetric part has the following expression
\be
\widetilde\nabla_\lambda {K^{\lambda}}_{\mu\nu}=4\pi G\,\imath\left(\overline\psi\gamma_{[\nu} \widetilde D_{\mu]}\psi-\widetilde {\overline D}_{[\mu}\overline\psi\gamma_{\nu]}\psi\right).
\ee{19}

%%%%%%%%%%%%%%%%%%%%%%%%%%%%%%%%%%%%%%%%%%%%%%%
\section{Field equations in a Robertson-Walker metric}
%%%%%%%%%%%%%%%%%%%%%%%%%%%%%%%%%%%%%%%%%%%%%%%

In a homogeneous and isotropic Universe described by  the spatially plane Robertson-Walker metric -- namely, $ds^2=dt^2-a(t)^2\left(dx^2+dy^2+dz^2\right)$ where $a(t)$ is the scale factor -- the antisymmetric part of Einstein's field equations (\ref{19}) becomes an identity so that it does not prescribe any restriction to the dynamics of the system, i.e., the evolution of the Universe  is ruled by the symmetric part given by equation (\ref{17}).

The Friedmann and acceleration equations obtained from the symmetric part of Einstein's field equations (\ref{17}) read
\be
\left({\dot a \over a}\right)^2={8\pi G\over3}\rho,\qquad
{\ddot a\over a}=-{4\pi G\over3}(\rho+3p).
\ee{20}
Above the total energy density $\rho$ and the pressure $p$ of the sources are given by
\ben\label{21}
\rho=m\overline\psi\psi+V-{3\pi G\over 2}\sigma^2,\qquad
p={\overline\psi\over2}{dV\over d\overline\psi}+{dV\over d\overline\psi}{\psi\over2}-V-{3\pi G\over 2}\sigma^2.\label{21}
\een

The Dirac equations (\ref{14}) in a spatially flat Robertson-Walker metric become
\ben\label{24}
\dot\psi&+&\frac{3}{2}\frac{\dot a}{a}\psi+\imath m\gamma^0\psi+\imath\gamma^0{dV\over d\overline\psi}=3{\pi G}\imath\gamma^0\left(\overline\psi\gamma_5\gamma^i\psi\right)\left(\gamma_5\gamma_i\psi\right),
\\\label{25}
\dot{\overline\psi}&+&\frac{3}{2}\frac{\dot a}{a}\overline\psi-\imath m\overline\psi\gamma^0-\imath{dV\over d\psi}\gamma^0
=-3{\pi G}\imath\left(\overline\psi\gamma_5\gamma^i\psi\right)\left(\overline\psi\gamma_5\gamma_i\right)\gamma^0.
\een

%%%%%%%%%%%%%%%%%%%%%%%%%%%%%%%%%%%%%%%%%%%%%%%
\section{Inflationary solution}
%%%%%%%%%%%%%%%%%%%%%%%%%%%%%%%%%%%%%%%%%%%%%%%

 In order to determine  the cosmic scale factor $a(t)$ and the components of the spinor field $\psi(t)=(\psi_1(t), \psi_2(t), \psi_3(t), \psi_4(t))^T$ one may use the acceleration equation (\ref{20})$_2$ and  the Dirac equation (\ref{24}). However, the search for exact solutions of this coupled system of five differential equations is a very hard job and   numerical solutions of this system of equations will be determined. In this work  a massive fermion field without a self-interaction potential is analyzed and in this case, the energy density and pressure  reduce to
 \ben\label{26}
\rho=m\overline\psi\psi-{3\pi G\over 2}\sigma^2,\qquad
p=-{3\pi G\over 2}\sigma^2,
\een
whereas the Dirac equation (\ref{24}) reads
\be
\dot\psi+\frac{3}{2}\frac{\dot a}{a}\psi+\imath m\gamma^0\psi
=3{\pi G}\imath\gamma^0\left(\overline\psi\gamma_5\gamma^i\psi\right)\left(\gamma_5\gamma_i\psi\right),
\ee{27}

For the search of the numerical solutions   the cosmological time and the fermion mass are written in terms of the new variables $\pi G t$ and $m/\pi G$, respectively.
Furthermore, in order to have initially a positive acceleration with a positive energy density one obtains from the Friedmann and acceleration equations (\ref{20}) the following restriction
\be
{3\over2}\sigma^2<{m\over \pi G}\overline\psi\psi<6\sigma^2.
\ee{28}
The initial conditions chosen in order to solve numerically the system of equations (\ref{20})$_2$ together with (\ref{26}) and (\ref{27}) were $a(0)=1$,  $\psi_1(0)=0.25, \psi_2(0)=0.15, \psi_3(0)=0.05$ and $\psi_4(0)=0.10$. Moreover, the initial condition for $\dot a(0)$ was obtained from the Friedmann equation (\ref{20})$_1$. In order to notice the role of the fermion mass five values were chosen, namely $m/\pi G=0.20;\, 0.25;\, 0.35;\, 0.40;\, 0.45$.

\begin{figure}
 \begin{center}
 \vskip0.5cm
 \includegraphics[height=5cm,width=7cm]{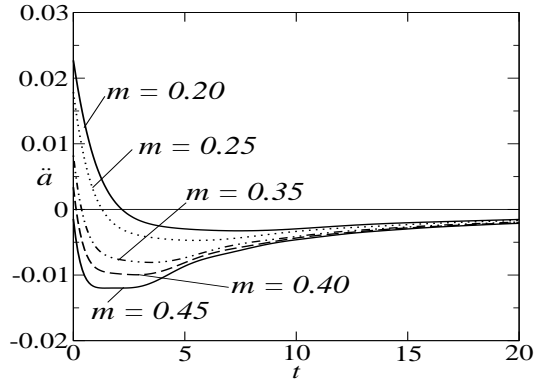}
   \caption{Acceleration field $\ddot a$ versus time $\pi G t$.}
  \label{fig.1}
 \end{center}
 \end{figure}

 In Figure \ref{fig.1} it is plotted the acceleration field $\ddot a$ as a function of time $\pi G t$ for  the five different masses.  One may infer from this figure that  there exist transitions from  accelerated   to  decelerated regimes for  $m/\pi G=0.20,\, 0.25,\, 0.35,\, 0.40$, whereas for  $m/\pi G=0.45$ only a decelerated regime occurs. Hence, the spin density of  a fermions field can promote accelerated regimes when its mass is not too large and the massive field is the responsible for the decelerated regime. Furthermore, the more massive field enters earlier in the decelerated regime and its deceleration is larger than the less massive field. In all cases  the deceleration tends asymptotically to zero for large times. It is noteworthy to call attention that massless fermion fields is ruled out from this analysis, since in this case the energy density of the fermionic field becomes negative.

\begin{figure}
 \begin{center}
 \vskip0.5cm
 \includegraphics[height=5cm,width=7cm]{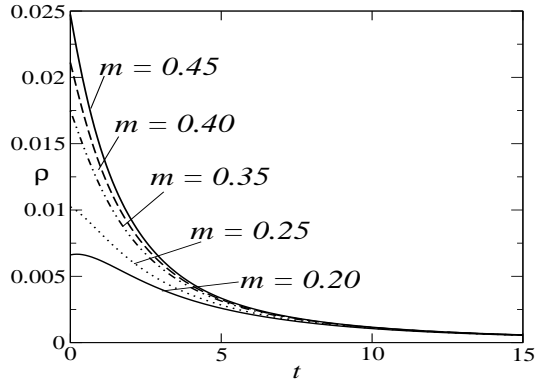}
   \caption{Energy densities  versus time $\pi G t$.}
   \label{fig.2}
 \end{center}
 \end{figure}

In Figure \ref{fig.2} it is represented  the energy densities as functions of time $\pi G t$. As was expected, the more massive fermionic field has the large energy density and decays more rapidly with time. All energy densities tend to a common value  going asymptotically to zero.

\section{Conclusions}
As a main conclusion one may infer that  within the context of the Einstein-Cartan theory a massive fermion field together with its spin density can generate an inflationary period and promote a transition to a matter field in the primordial Universe. The  acceleration is due to the spin density but it decays rapidly and the term associated with the mass of the fermion field contributes to a  deceleration of  the Universe. A similar behavior can be found if one considers a non massive fermion field with a self-interaction potential given by $V=V_0\overline\psi\psi$. Hence the inflationary and the transition from the accelerated to a decelerated regime can be achieved by only one field, namely, a fermion field  and  its associated spin density.

\section*{Acknowledgments}
GMK acknowledges the support by
Conselho Nacional de Desenvolvimento Cient\'\i fico e Tecnol\'ogico (CNPq).

\end{document}